\documentclass[%
 reprint,
superscriptaddress,
 amsmath,amssymb,
 aps,
pre,
floatfix,
showpacs
]{revtex4-1}
\usepackage[normalem]{ulem}
\usepackage{graphicx}
\usepackage{dcolumn}
\usepackage{bm}
\usepackage{hyperref}

\hypersetup{colorlinks=true,citecolor=blue,linkcolor=cyan,urlcolor=blue,filecolor= green, breaklinks=true}

\begin{document}

\title{Quantum correlated cluster mean-field theory applied to the transverse Ising model}

\author{F. M. Zimmer}
 \email{fabiozimmer@gmail.com}
 \affiliation{Departamento de F\'isica, Centro de Ci\^encias Naturais e Exatas, Universidade Federal de Santa Maria, Avenida Roraima 1000, 97105-900, Santa Maria, RS, Brazil}
 
\author{M. Schmidt}
 \email{mateusing85@gmail.com}
 \affiliation{Departamento de F\'isica, Centro de Ci\^encias Naturais e Exatas, Universidade Federal de Santa Maria, Avenida Roraima 1000, 97105-900, Santa Maria, RS, Brazil}

\author{Jonas Maziero}
 \email{jonas.maziero@ufsm.br}
 \affiliation{Departamento de F\'isica, Centro de Ci\^encias Naturais e Exatas, Universidade Federal de Santa Maria, Avenida Roraima 1000, 97105-900, Santa Maria, RS, Brazil}
 \affiliation{Instituto de F\'isica, Facultad de Ingenier\'ia, Universidad de la Rep\'ublica, J. Herrera y Reissig 565, 11300, Montevideo, Uruguay}


\begin{abstract}
Mean-field theory (MFT) is one of the main available tools for analytical calculations entailed in investigations regarding many-body systems. Recently, there have been an urge of interest in ameliorating this kind of method, mainly with the aim of incorporating geometric and correlation properties of these systems. The correlated cluster MFT (CCMFT) is an improvement that succeeded quite well in doing that for classical spin systems. Nevertheless, even the CCMFT presents some deficiencies when applied to quantum systems. In this article, we address this issue by proposing the quantum CCMFT (QCCMFT), which, in contrast to its former approach, uses general quantum states in its self-consistent mean-field equations. We apply the introduced QCCMFT to the transverse Ising model in honeycomb, square, and simple cubic lattices and obtain fairly good results both for the Curie temperature of thermal phase transition and for the critical field of quantum phase transition. 
Actually, our results match those obtained via exact solutions, series expansions or Monte Carlo simulations.
\end{abstract}


\maketitle


\section{Introduction}
\label{introduction}

 The physics of many-body systems (MBS) is relevant for investigations in several research areas, such as condensed matter, complex systems, artificial intelligence, and quantum information science \cite{Anderson_MiD, Kadanoff_MFT, Dorogovtsev2008, Petruccione_NN, Buice2013, DelasCuevas1180}. Magnetic models have been particularly useful for studying phenomena related to phase transitions in MBS. Notwithstanding, there are few such Hamiltonians with known exact solution \cite{Baxter1989, Sutherland2004}. For instance, the Ising model in the presence of a transverse magnetic field is the simplest model presenting a zero temperature phase transition driven by quantum fluctuations, but only the one dimensional case has been given an exact solution \cite{Sachdev_book}. Therefore, there is an evident demand for approximative approaches which are able to provide an informative and computationally efficient description of the main features of MBS.
 
 In this direction, several methods have been applied for analyzing MBS. For instance, the Monte Carlo (MC) method \cite{Binder_MCM} is frequently utilized to study classical problems, though its application to quantum systems may be intricate. On the other hand, we can use MC simulations to describe, for instance, the $d$-dimensional quantum Ising model (QIM) by exploring  its $(d+1)$-dimensional classical counterpart \cite{suzuki,CMC_Deng}. The critical properties of the QIM can also be investigated via others techniques such as, for example, the normalization group \cite{RG}, cluster variational method \cite{PhysRev.81.988,PhysRevB.27.6884}, matrix product states and projected entangled pair states \cite{Orus2014}, series expansions \cite{ser_exp_Quantum_Ising,ser_exp_Quantum_Ising3D}, 
and mean-field theory (MFT) \cite{Kadanoff_MFT}. Because of its amenability for analytical calculations, MFT has been frequently adopted as the starting point for investigations regarding magnetic systems, with the interacting many-body Hamiltonian being approximated in order to obtain an effective single-particle one. The MFT is capable of correctly describe the qualitative behavior of many systems, but it neglects some geometric features of the spin lattice and most correlations among its constituent particles \cite{Mattis1979, Zhuravlev2005}. As a matter of fact, the results obtained with MFT are highly dependent on the lattice coordination number $z$. Nevertheless, MFT's versatility and simplicity stimulate its development targeting to improve its quantitative predictions by including correlation and geometry effects. 
 
 A straightforward betterment to the standard MFT is obtained by considering spin clusters instead of a single spin.  
In the so called cluster MFT (CMFT), the intracluster interactions are calculated exactly while the intercluster ones are evaluated following a mean-field variational procedure \cite{PSS_Ferreira, PhysRevB.43.6181, variational_CMF}. This means that the correlations neglected within the MFT could be gradually recovered as the cluster size increases. However, similarly to exact diagonalization, the CMFT is limited by the needed computational ability to evaluate large clusters
\cite{variational_CMF}. Methods based on self-consistent conditions have also been developed in a mean-field framework. For example, the recently proposed entanglement mean-field theory (EMFT) is based on a two-body self consistent equation \cite{SenDe2011, SenDe2012, aditi, EMFT_Sen_De_Calssical}.  The EMFT applied to the quantum Ising model improves the MFT results for the critical values of temperature and transverse field, but it is still highly dependent on $z$, not including other lattice geometric features.
 

Another important step forward in the development of ameliorated MFTs was given by the correlated cluster mean-field theory (CCMFT) \cite{Yamamoto}.
 This technique extends the single-site correlated molecular-field theory, introduced in Ref. \cite{Wysin2000}, to a cluster approach, in which self-consistent mean-field equations dependent on the cluster spin configurations are considered. The CCMFT estimates with good accuracy the critical temperatures when applied to the classical Ising model for different lattice geometries  \cite{Yamamoto}. In addition, the CCMFT improves the behavior of short-range correlations as compared to other MFT-like calculations. It is important to remark that the CCMFT deals with small clusters, avoiding thus the computing time overload of the CMFT, and is free from the strong dependence with $z$ inherent to MFT and EMFT. In this sense, the CCMFT goes beyond its cousins MFTs by taking into account more lattice geometric features. As a consequence, the CCMFT leads to different critical temperatures for systems that have the same coordination number, as e.g. the kagome and square or the triangular and simple cubic lattices \cite{Yamamoto, Schmidt2015416}. Besides, the CCMFT has been successfully adopted to analyze several classical  spin-1/2 interacting systems \cite{YamamotoJPCF_2010, Chvoj2010}, including disordered \cite{ZimmerSchmidt2014} and geometrically frustrated \cite{Schmidt2015416} lattices. However, up to now, it has not been applied to the QIM, and filling this gap is one of our main contributions in this article. Besides, it is worthwhile mentioning the fact that, when applied to the QIM, the CCMFT leads to inconsistent results with the appearance of a discontinuity in the magnetization
 when the quantum fluctuations are strong enough. So, it is necessary to adapt this interesting technique for dealing with quantum systems in a thorough manner.

 Motivated by these observations, we propose a quantum version of the CCMFT (QCCMF) that allows us to describe very well the phase diagram of the QIM for different lattice geometries. Like in the CCMFT, in the QCCMFT the spin lattice is divided into clusters with the intercluster interactions being replaced by mean-fields that depend on the possible states of the cluster spins. The main difference between the two methods is that while the CCMFT restricts these states to the classical quantization axes, the QCCMFT admits more general quantum states. Here, this set of quantum states is self-consistently computed by using the imaginary time spin self-interaction. The resulting effective cluster model is then solved by exact diagonalization for different lattice geometries. Magnetization and spin correlations are evaluated for the classical and quantum regimes. The critical values of temperature and transverse field are obtained with very good accuracy when compared with other MFTs. In fact, these results can be matched to those obtained via state of the art MC simulations.

The remainder of this article is structured as follows. In the next section (Sec. \ref{model}),  we present the QCCMFT applied to the QIM in the honeycomb (Sec. \ref{honeycomb}), square (Sec. \ref{square}), and simple cubic (Sec. \ref{cubic}) lattices. Sec. \ref{results} is dedicated to describe and discuss the numerical results for magnetization, correlation functions, and phase diagrams of temperature versus transverse field. We summarize the article and make some concluding remarks in Sec. \ref{conclusion}. The calculation of the imaginary time spin self-interaction is provided in the Appendix \ref{sec_appendix}.


\section{The quantum correlated cluster mean field theory}
\label{model}

Let us begin this section by recalling that the Hamiltonian of the quantum Ising model can be written in the following manner:
\begin{equation}
 H= - J\sum_{(i,j)}^{N} \sigma_{i}^{z}\sigma_{j}^{z}-\Gamma \sum_{i=1}^{N}\sigma_{i}^{x},
 \label{ham}
 \end{equation}
where $\sigma_{i}^{z}$ and $\sigma_{i}^{x}$ are the Pauli spin operators acting on the Hilbert space of the $i$-th spin ($\mathcal{H}_{i}$), $J$ sets the energy unit for exchange interactions, $\Gamma$ is the transverse magnetic field, and $(i,j)$ indicates that the first sum is made over all the nearest neighbors in a given lattice with $N$ sites. 

\begin{figure*}
  \includegraphics[width=2.\columnwidth]{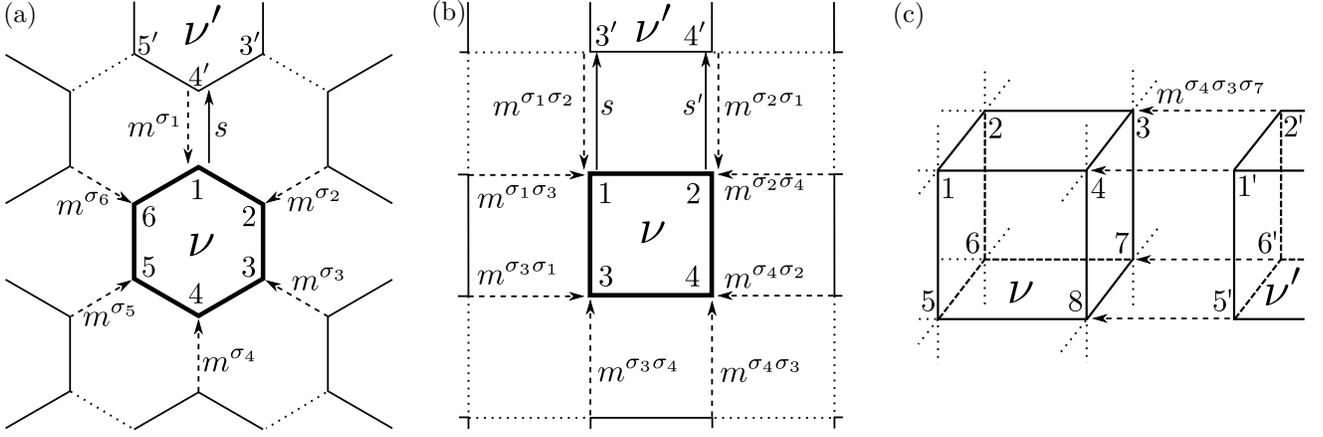}
 \caption{Schematic representation of the (a) honeycomb, (b) square, and (c) simple cubic lattices divided into clusters of spins. The dashed arrows indicate mean-fields. The solid arrows correspond to the expected values used to evaluate the mean-fields.  }
 \label{fig1}
\end{figure*}

In the CCMFT, the themodynamic limit ($N\rightarrow \infty$) of the model is approximated by dividing the spin lattice into identical clusters with $n_s$ spins each and in such a way that the resulting set of clusters follows the original lattice symmetry. The intracluster interactions are fully preserved while the intercluster ones are approximated using correlated-effective fields. 
This procedure results in the following effective single-cluster quantum model:
\begin{equation}
 H_{\nu}^{eff}=  H_{\nu} + H_{inter},
\label{H_eff_central}
\end{equation}
with
\begin{equation}
 H_{\nu} = - J\sum_{(i,j)}^{n_{s}} \sigma_{i}^{z}\sigma_{j}^{z} -\Gamma \sum_{i=1}^{n_s}\sigma_{i}^{x}
\end{equation}
representing the intracluster interactions and the sums being made over the $n_s$ sites of a cluster $\nu$, see Fig. \ref{fig1}.
By its turn, the intercluster interactions are approximated by effective fields that act on the cluster boundary sites, that is to say,
\begin{equation}
H_{inter}=- J\sum_{i=1}^{n_s}\sigma_{i}^{z}h^{eff}_{i}.
\label{Hint}
\end{equation}
The effective field $h^{eff}_{i}$ depends on the states of the spin $i$ and of its neighbors which belong to the same cluster boundary.
Therefore, different lattice geometries can have different numbers of neighbors between clusters and consequently different effective fields (see Fig. \ref{fig1}). In the next sub-sections, we exemplify this procedure by regarding explicitly three types of lattices: honeycomb, square, and simple cubic.

\subsection{Honeycomb Lattice}
\label{honeycomb}

 The honeycomb lattice can be divided into topologically equivalent clusters of six sites each ($n_s=6$), and in such a way that there is only one pair of interacting spins between neighboring clusters (see Fig. \ref{fig1}(a)). This intercluster interaction is replaced by mean fields $m^{+}$ and $m^{-}$ acting on the spins of the considered cluster $\nu$ \cite{Yamamoto}. And, as in the CCMFT, these mean fields are determined by the ``states'' of the site in which they act on ($|\uparrow\rangle$ or $|\downarrow\rangle$). As a result,  $H_{inter}$ can be recast as:

\begin{equation}
H_{inter} = -J\sum_{i=1}^{n_s}\left(\frac{\mathbb{I}_{i}+\sigma_{i}^{z}}{2}m^{+} - \frac{\mathbb{I}_{i}-\sigma_{i}^{z}}{2}m^{-}\right),
\label{heff_hex}
\end{equation}
with $\mathbb{I}_{i}$ being the indentity operator in $\mathcal{H}_{i}$.

Now let's regard the computation of $m^{+}$ and $m^{-}$. In order to incorporate the quantum nature of this kind of system, in the QCCMFT we do not restrict the ``states'' used to compute the mean fields to be in the classical quantization axes. To instantiate how we do that, let us start considering the cluster $\nu'$ in Fig. \ref{fig1}(a), whose site $4'$ is first neighbor to site $1$ in cluster $\nu$. As for $\nu$, effective fields, $h_i^{eff}$, act on the border spins of cluster $\nu'$, with exception to the site $4'$. In the QCCMFT, the spin at this site interacts with the expected value of the magnetic moment of site $1$ for two general orthogonal states:
\begin{eqnarray}
|\Phi^{+}\rangle & = & \cos(\theta/2)|\uparrow\rangle + e^{i\phi}\sin(\theta/2)|\downarrow\rangle, \\
|\Phi^{-}\rangle & = & \sin(\theta/2)|\uparrow \rangle -e^{i\phi}\cos(\theta/2)|\downarrow\rangle,
\end{eqnarray}
where the angles $\theta\in[0,\pi]$ and $\phi\in[0,2\pi)$ have to be determined self-consistently. For the QIM there exists an ``easy'' axes for the exchange interaction, the $z$ direction. So, once only the $z$ component of the magnetic moment of the spin 1 is expected to be relevant for its interaction with spin $4'$, and
\begin{equation}
 \langle \Phi^{\pm}| \sigma_1^{z} | \Phi^{\pm}\rangle = \pm\cos\theta,
\end{equation}
we shall use $\phi=0$. We observe that if we set $\theta = 0$, then the QCCMFT is equivalent to the CCMFT.

Thus, the self-consistent mean-field equations read
\begin{equation}
 m^{s}=\mbox{Tr}\left(\sigma_{4'}\rho_{\beta}(H^{eff}_{\nu'}(s))\right),
 \label{ms}
\end{equation}
with
\begin{equation}
 \rho_{\beta}(H^{eff}_{\nu'}(s)) = \frac{\exp(-\beta H^{eff}_{\nu'}(s))}{\mbox{Tr}\exp(-\beta H^{eff}_{\nu'}(s))}
\end{equation}
being the Gibbs thermal state for the Hamiltonian
\begin{equation}
H^{eff}_{\nu'}(s) = H_{\nu'} -J\sum_{i\ne 4'}\sigma_{i}^{z}h_{i}^{eff} -J\sigma_{4'}^{z}s\cos\theta,
\end{equation}
where the sums are performed over the sites of cluster $\nu'$, $s=\pm 1$, and $\beta=1/T$ with $T$ being the temperature (we set the Boltzmann constant to unit).

Before solving the equations for $m^{s}$, we have to input the value of $\theta$. In this article, our ansatz is to provide this quantity 
via the imaginary time spin self-interaction. As discussed and motivated in Appendix \ref{sec_appendix}, we obtain $\theta$ using the following relation
\begin{equation}
 \bar{q}=\left\langle\frac{1}{\beta}\int_{0}^{\beta}d\tau  \sigma_{i}^{z}\sigma_{i}^{z}(\tau) \right\rangle_{\rho_{\beta}(H_{\nu}^{eff})} = \cos^{2}\theta\label{ss}.
\end{equation}
With this, Eqs. (\ref{ms}) and  (\ref{ss}) are solved self-consistently to obtain the mean fields $m^{+}$ and $m^{-}$. These fields are finally used in Eq. (\ref{heff_hex}) to yield the effective model (\ref{H_eff_central}) for the honeycomb lattice, from which one can compute all physical quantities of interest. The results for this lattice are presented in the next section.

\subsection{Square Lattice}
\label{square}

This lattice is divided into clusters with $n_s=4$ sites each, and there are, for each site, two spin interactions between neighboring clusters (see Fig. \ref{fig1}(b)). In the CCMFT, the mean field due to a site in the cluster $\nu'$ is strongly dependent on the spin states of its first and second neighbors in the cluster $\nu$.
Analogously to the previous sub-section, in the standard-Ising basis, these spin states are $|\uparrow\uparrow\rangle$, $|\uparrow\downarrow\rangle$, $|\downarrow\uparrow\rangle$, and  $|\downarrow\downarrow\rangle$, which are associated, respectively, with the mean fields $m^{++}$, $m^{+-}$, $m^{-+}$, and $m^{--}$.
Following the CCMFT procedure \cite{Yamamoto}, the effective single cluster model  (\ref{H_eff_central}) can be expressed with $H_{inter}$ written as
\begin{eqnarray}
H_{inter} & = & -J\sum_{i,k}\left(\frac{\mathbb{I}_{i}+\sigma_{i}^{z}}{2}\frac{\mathbb{I}_{k}+\sigma^{z}_{k}}{2}m^{++}
\nonumber \right. \\ 
&& \hspace{1.3cm} - \frac{\mathbb{I}_{i}+\sigma_{i}^{z}}{2}\frac{\mathbb{I}_{k}-\sigma^{z}_{k}}{2}m^{+-} \nonumber \\ 
&& \hspace{1.3cm} - \frac{\mathbb{I}_{i}-\sigma_{i}^{z}}{2}\frac{\mathbb{I}_{k}+\sigma^{z}_{k}}{2}m^{-+} \nonumber \\
&& \left. \hspace{1.3cm} + \frac{\mathbb{I}_{i}-\sigma_{i}^{z}}{2}\frac{\mathbb{I}_{k}-\sigma^{z}_{k}}{2}m^{--} \right),
\label{eqsquare0}
\end{eqnarray}
where, for the square lattice, the pairs of indexes are $(i,k)=(1,2)\mbox{, }(2,4)\mbox{, }(4,3)\mbox{, }(3,1)$ and its permutations.

The mean fields 
are obtained by considering the nearby connected clusters $\nu'$. 
For instance, $m^{ss'}$ is determined from the expected value of $\sigma_{3'}^{z}$ when the intercluster interaction for $\nu'$ is given by effective fields as in the equations below, except for the connected sites $3'$ and $4'$.  
These sites interact with the expected values of magnetic moments of sites 1 and 2 of cluster $\nu$ on the state $|\Psi^{ss'}\rangle.$
Explicitly, the mean fields are computed by solving self-consistently the set of equations:
\begin{equation}
m^{ss'}=\mbox{Tr}\left(\sigma_{3'}^{z}\rho_{\beta}(H_{\nu'}^{eff}(ss'))\right),
\end{equation}
where $3'$ is a site of cluster $\nu'$ that is neighbor of sites 1 and 2 of cluster $\nu$.
The effective Hamiltonian $H_{\nu'}^{eff}(ss')$ of cluster $\nu'$ is expressed by 
\begin{equation}
 H_{\nu'}^{eff}(ss')=H_{\nu'} + H^{eff}_{inter}(ss'),  
\end{equation}
with
\begin{eqnarray}
 H^{eff}_{inter}(ss') & = & -\frac{J}{4} \sum\limits_{\substack{    i \\
     \{i,k\}\ne \{3',4'\}, \{4',3'\}}}^{n_s} \\
&& [\sigma_{i}^{z} (m^{++}+m^{+-}+m^{-+}+m^{--}) \nonumber \\
&& +\sigma_{i}^{z}\sigma_{k}^{z} (m^{++}-m^{+-}+m^{-+}-m^{--}) \nonumber \\
&&  +\sigma_{k}^{z} (m^{++}-m^{+-}-m^{-+}+m^{--})] \nonumber \\
&& -J\langle\Psi^{ss'}|\sigma_{1}^{z}|\Psi^{ss'}\rangle \sigma_{3'}^{z} -J \langle\Psi^{ss'}|\sigma_{2}^{z}|\Psi^{ss'}\rangle \sigma_{4'}^{z}. \nonumber 
\end{eqnarray}
In particular, our ansatz to the states $|\Psi^{ss'}\rangle$ is: 
\begin{eqnarray}
|\Psi^{++}\rangle & = & (\cos (\theta/2)|\uparrow\rangle +\sin (\theta/2) |\downarrow\rangle)|\uparrow\rangle, \\
|\Psi^{+-}\rangle & = & (\cos (\theta/2)|\uparrow\rangle +\sin (\theta/2) |\downarrow\rangle)|\downarrow\rangle, \\
|\Psi^{-+}\rangle & = & (\cos (\theta/2)|\downarrow\rangle -\sin (\theta/2) |\uparrow\rangle)|\uparrow\rangle, \\
|\Psi^{--}\rangle & = & (\cos (\theta/2)|\downarrow\rangle -\sin (\theta/2) |\uparrow\rangle)|\downarrow\rangle. 
\end{eqnarray}
These states result in the following expected values:
\begin{equation}
\langle\Psi^{ss'}|\sigma_{1}^{z}|\Psi^{ss'}\rangle = s \cos(\theta) \mbox{ and } \langle\Psi^{ss'}|\sigma_{2}^{z}|\Psi^{ss'}\rangle = s'.
\end{equation}
Here the imaginary time spin self-interaction of cluster $\nu$ is used once more to obtain $\theta= \cos^{-1}\sqrt{\bar{q}}$ (see appendix \ref{sec_appendix}), where $\bar{q}$ is evaluated considering the square lattice effective model (\ref{H_eff_central}) with $H_{inter}$ given by Eq. (\ref{eqsquare0}).

\subsection{Simple Cubic Lattice}
\label{cubic}

For the three-dimensional case, we consider a simple cubic lattice. For this system, the calculation is performed based on eight-site cubic clusters, as illustrated in Fig. \ref{fig1}(c). There are four sites connected between neighbor cluster faces, what introduces sixteen mean fields in the original CCMFT. However, we assume an approach that takes into account only three connected sites per face. For instance, we consider the states of site $i$ and its nearest neighbors of the same cluster face $\nu$ to obtain the mean fields that act on the spin of site $i$ \cite{ZimmerSchmidt2014}. In this case, the number of mean fields is decreased to eight: $m^{+++}$, $m^{++-}$, $m^{+-+}$, $m^{-++}$, $m^{+--}$, $m^{-+-}$, $m^{--+}$, and $m^{---}$.   
We can also explore the symmetries $m^{++-}=m^{-++}$ and $m^{--+}=m^{+--}$ to evaluate only six mean fields, reducing thus the numerical cost in this case.
The procedure is a straightforward extension of that in the case of the square lattice. However, here we consider the following orthogonal set of states $|\psi^{s'ss''}\rangle$ instead of $|\Psi^{ss'}\rangle$, with $|\psi^{+++}\rangle= \cos(\theta/2)|\uparrow\uparrow\uparrow\rangle + \sin(\theta/2)|\uparrow\downarrow\uparrow\rangle$, $\cdots$, $|\psi^{---}\rangle= \cos(\theta/2)|\downarrow\downarrow\downarrow\rangle - \sin(\theta/2)|\downarrow \uparrow\downarrow\rangle$.


\section{Numerical results and discussion}
\label{results}

The mean fields are determined numerically from the effective hamiltonian of cluster $\nu'$  in a self-consistent routine by using an exact diagonalization method. In this framework, the lattice geometric features are addressed by the cluster structure.
The effective hamiltonian of $\nu'$ also depends on $\theta$, which is evaluated from the imaginary time spin self-interaction $\bar{q}$ of cluster $\nu$. 
In other words, this routine requires only two clusters: the central cluster  $\nu$ and one of its nearby connected $\nu^{'}$. The cluster $\nu^{'}$ is used to compute the mean fields and from $\nu$ the thermodynamic quantities can be derived.  Therefore, this
procedure allows us to get an effective single cluster model, Eq. (\ref{H_eff_central}), that can be used to obtain the observables of interest  as magnetization,  
\begin{equation}
m^{\alpha}=\mbox{Tr}\left(\sigma^{\alpha}_{i}\rho_{\beta}(H_{\nu}^{eff})\right),
 \end{equation}
and spin correlation,
\begin{equation}
C_{\alpha \alpha}=\mbox{Tr}\left(\sigma^{\alpha}_{i}\sigma^{\alpha}_{j}\rho_{\beta}(H_{\nu}^{eff}) \right),
 \end{equation}
 where $\alpha=x,z$ refers to the component of the Pauli spin operator.

\begin{figure}
 \includegraphics[width=1.0\columnwidth]{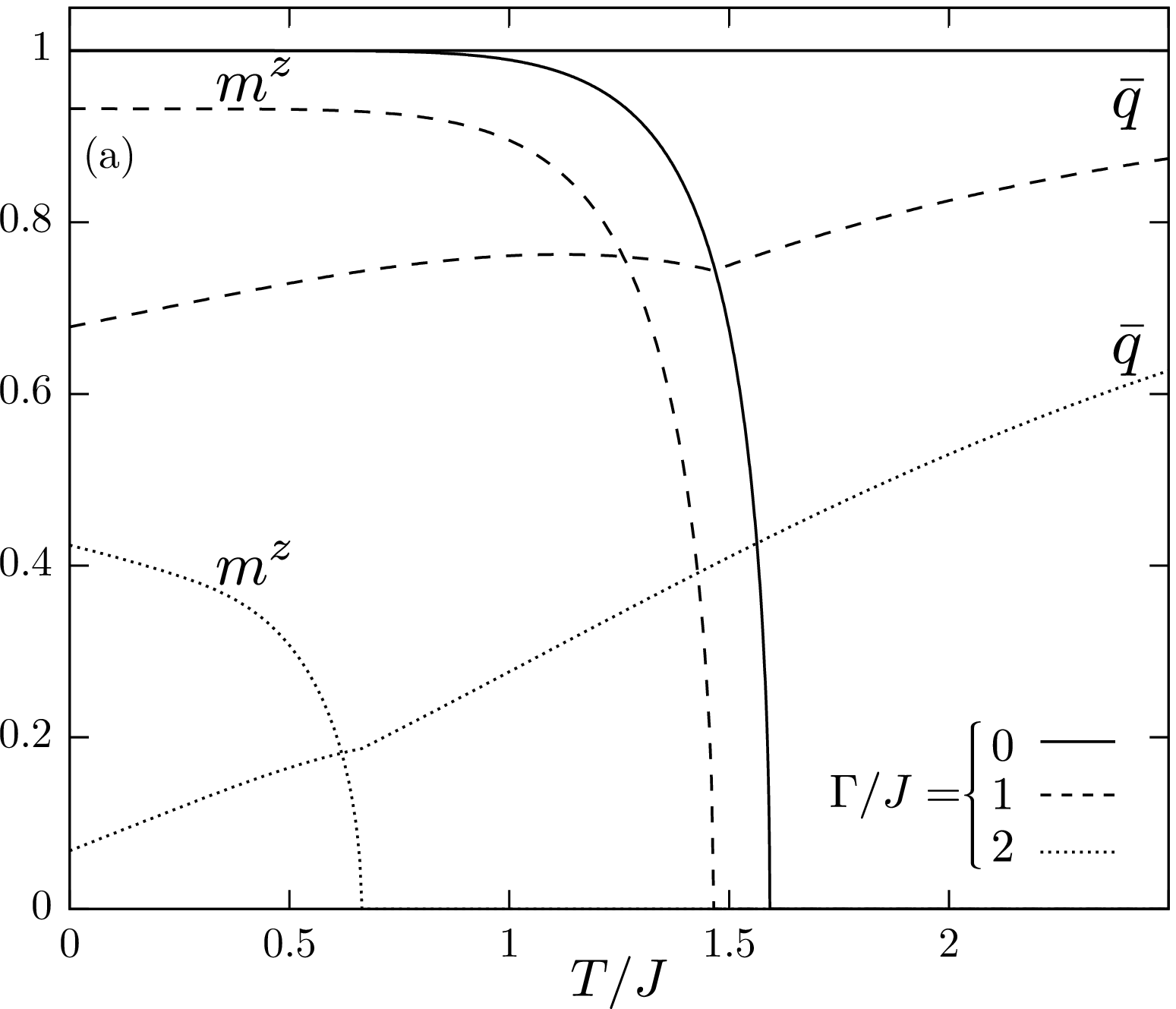}
 
 \vspace{0.4cm} 
 
 \includegraphics[width=1.\columnwidth]{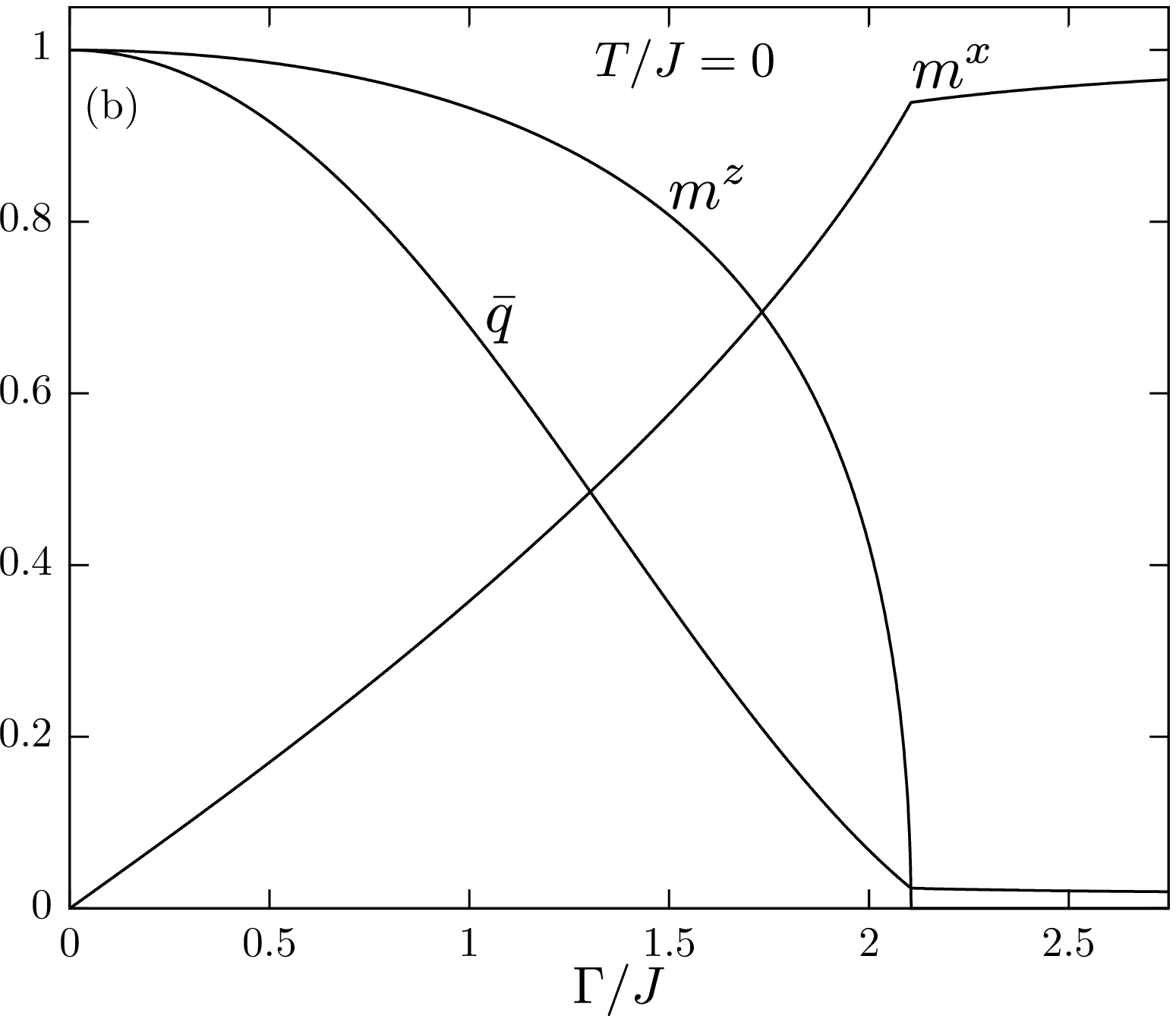}
 \caption{ Results for the honeycomb lattice: (a) Temperature dependence of $m^z$ and $\bar{q}$ for different intensities of $\Gamma/J$. (b) $m^z$, $m^x$ and $\bar{q}$ as a function of $\Gamma/J$ for $T=0$.  }
 \label{fig_hex}
\end{figure}

 The honeycomb lattice results can be analysed in Fig. \ref{fig_hex}. For $\Gamma=0$, when the temperature decreases, the system presents spontaneous magnetization $m^z$ below the critical temperature $T_c$. In addition, the absence of quantum fluctuations leads to a spin self-interaction independent of temperature. As a consequence, $\bar{q}=1$ and the classical CCMFT results are recovered with $T_c=1.592 J$ \cite{Yamamoto}. 
 When $\Gamma > 0$, $m^z$ and $T_c$ are gradually decreased as the quantum fluctuations increases (see Fig. \ref{fig_hex}(a)). Moreover, $\bar{q}$ is very sensitive to the presence of quantum fluctuations. Thus, $\bar{q}$ becomes temperature dependent for $\Gamma>0$. It exhibits a clear mark at $T_c$, going towards one at higher temperatures, where the thermal fluctuations dominate.

\begin{figure}
 \includegraphics[width=1.\columnwidth]{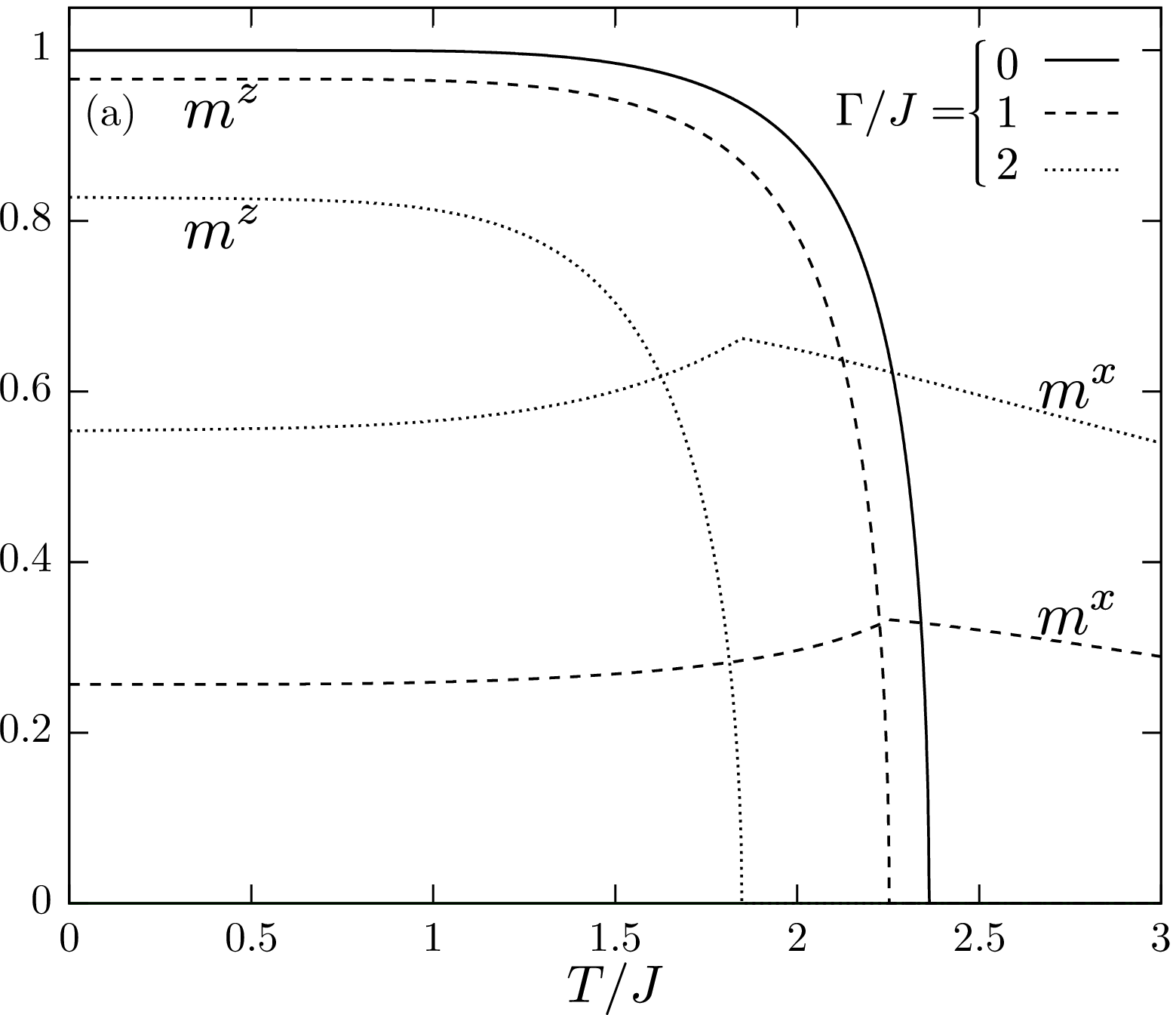}
 
 \vspace{0.4cm} 

 \includegraphics[width=1.\columnwidth]{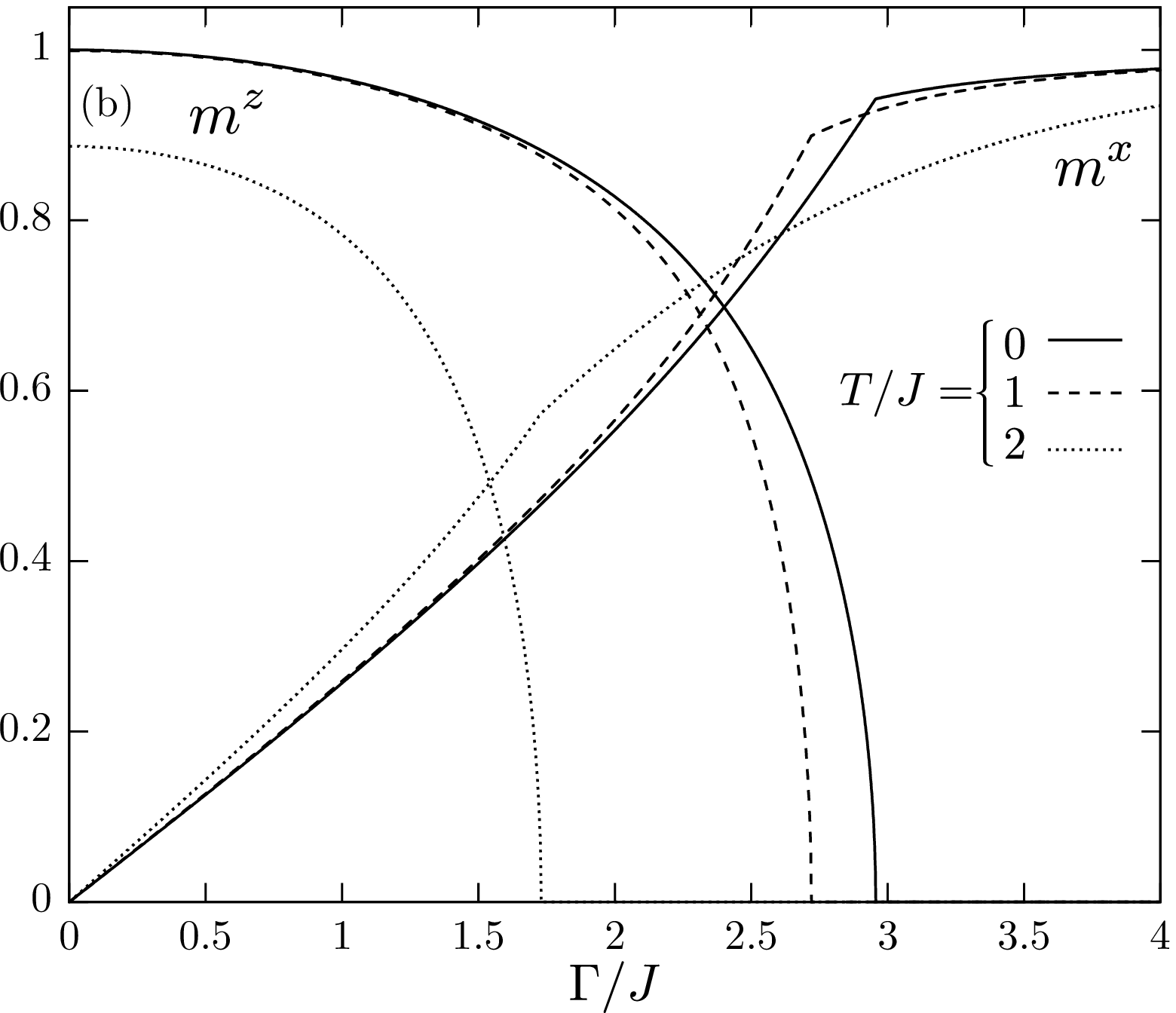}
 \caption{Results for the square lattice: (a) Temperature dependence of $m^z$, and $m^{x}$ for different intensities of $\Gamma/J$. (b) $m^z$ and $m^{x}$ as a function of $\Gamma/J$ for some values of $T/J$.  }
 \label{fig2}
\end{figure}

The effects of quantum fluctuations in the ground-state can be analysed in Fig. \ref{fig_hex}(b). The increase of $\Gamma$ leads the system to a quantum critical point (QCP) at $\Gamma_c=2.105 J$, where $m^z$ becomes zero. 
 The  magnetization $m^x$ arises when $\Gamma$ increases, exhibiting a mark at $\Gamma_c$. For $\Gamma>\Gamma_c$, $m^x$ increases monotonically towards one. Furthermore, $\bar{q}$ decreases faster within the ferromagnetic (FE) order than in the quantum paramagnetic (PM) phase, in which $\bar{q} \to 0$  when $\Gamma\rightarrow\infty$.

The magnetization results for the square lattice are presented in Fig. \ref{fig2}. The $m^{z}$ for $\Gamma=0$ recovers the classical CCMFT with the critical temperature $T_c(\Gamma=0)=2.362J$ and $m^{x}=0$.
As the transverse field increases, the magnetization $m^{z}$ decreases and  $m^{x}$ arises showing a cusp at $T_c$ (see Fig. \ref{fig2}(a)).
The effect of $\Gamma$ on the magnetizations can also be analysed in Fig. \ref{fig2}(b), that exhibits a QCP at $\Gamma_c=2.956 J$. 
In particular, $\bar{q}$ presents the same qualitative behavior as the one observed for the honeycomb lattice.

The study of magnetizations is also done for the simple cubic lattice, with the results being qualitatively equivalent to those in Fig. \ref{fig2}. However, the quantum critical point is  obtained at $\Gamma_c=5.198J$ and the critical temperature for the classical limit is found to be $T_c(\Gamma=0)=4.763J$.  This $T_c(\Gamma=0)$ value is very close to that obtained from the original CCMFT with 16 mean fields: $T_c(\Gamma=0)=4.753J$ \cite{Yamamoto}. It is important to remark that in our approach the problem is simplified by considering only six mean fields. 

\begin{figure} 
 \center{\includegraphics[width=1.\columnwidth]{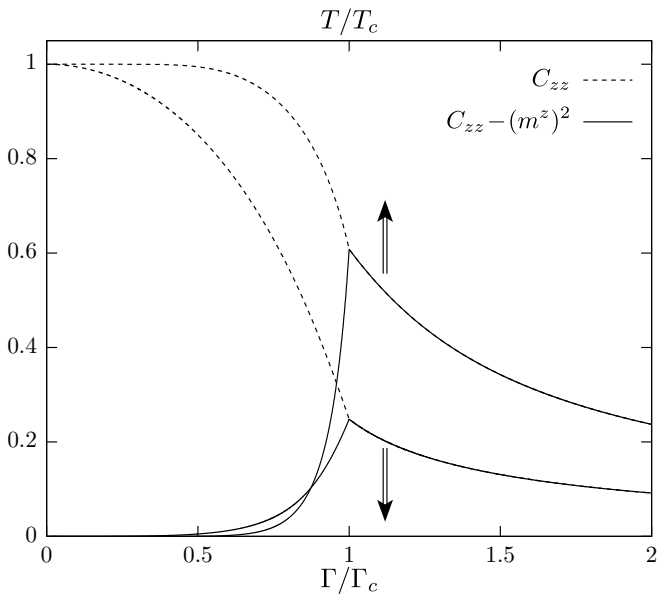}}
 \caption{Correlation as a function of reduced temperature (transverse field) at $\Gamma=0$ ($T=0$) for the square lattice. The fluctuations are also exhibited.}\label{corr}
\end{figure}
 Figure \ref{corr} shows the longitudinal correlation functions, $C_{zz}$ and the connected one $C_{zz}-(m^{z})^2$, between first neighbors in a square lattice. For the case of $\Gamma=0$ (upper curves), the thermal fluctuations decrease $C_{zz}$ and the neighbor spins become totally uncorrelated only at higher temperatures. At zero-temperature (lower curves), the quantum fluctuations lead the connected correlation function to a maximum at $\Gamma_c$. This maximum is lower than that obtained for the thermal fluctuations (classical) case.  It is noted that, differently from the MFT and EMFT, the QCCMFT leads to correlation functions different from zero in the classical or quantum paramagnetic regimes close to the phase transition. 
 On the other hand, we observe that, like in other mean-field approaches, the critical exponents obtained with the QCCMFT are the classical ones.

\begin{figure} 
 \center{\includegraphics[width=1.\columnwidth]{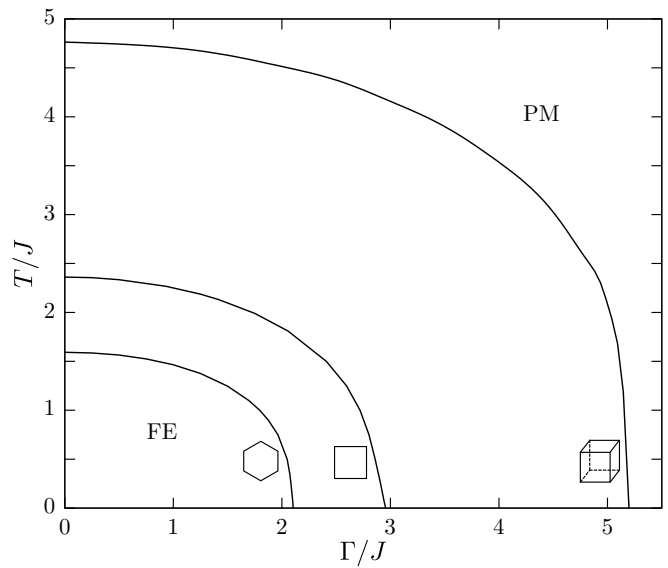}}
 \caption{Phase diagrams of temperature versus transverse field for the honeycomb, square, and cubic lattices.}\label{phase}
\end{figure}

The behavior of  the critical temperature as a function of $\Gamma$ is shown in Fig. \ref{phase} for the three lattices studied. 
The FE-PM phase transitions can occur by changes in thermal or quantum fluctuations and the phase boundaries are always continuous.
The critical temperatures diminish towards QCPs as $\Gamma$ increases. The QCP location not only depends on the dimension of the system, but also on the geometry of the lattice.      
These phase diagrams are in qualitative agreement with other mean field techniques. Furthermore, they also exhibit a very good quantitative agreement with the results presented in Fig. 3.5 of Ref. \cite{suzuki}, which were obtained via Monte Carlo simulations (MCS) and series expansions applied to the quantum Ising model.   

Table \ref{table} shows a comparison between the critical temperatures and the QCPs obtained from several methods applied to the quantum Ising model. 
 For instance, the EMFT and the effective-field theory (EFT), which is a single-site approximation based on the differential operator technique,  provide better results \cite{EFT_Sq_Sc, EFT_honey,EF_honey_Sq} for $T_c$ and $\Gamma_c$ as compared to the standard MFT. However, the MFT, EFT, and EMFT lead to results that are uniquely dependent on the lattice coordination number, neglecting other geometric features. In fact, $T_c=\Gamma_c$ in the MFT and EMFT \cite{aditi}.
On the other hand, the locations of these critical points from the QCCMFT method do not present a strong dependence on $z$.
In addition, the QCPs evaluated from the QCCMF approach are extremely close to those obtained from the best results of cluster Monte Carlo simulations \cite{CMC_Deng}. By comparing the QCCMFT results with the MCS \cite{MC_sc, CMC_Deng}  and exact results \cite{PhysRev.65.117}, the $\Gamma_c$ differences are less than $3\mbox{ }\%$, while the $T_c$ is overestimated by less than $6\mbox{ }\%$ for the considered lattices.

\begin{table}[ht]
  \centering
  \caption{ Comparison of the critical transverse fields and temperatures (in units of $J$) for the quantum Ising model on three classes of lattices and obtained using different methods.}
\begin{tabular}   {p{0.17\linewidth}p{0.05\linewidth}p{0.07\linewidth} p{0.10\linewidth} p{0.11\linewidth} p{0.18\linewidth} p{0.18\linewidth}p{0.0001\linewidth}}
    \hline
    Lattice      &       & \centering MFT  & \centering EFT &  \centering  EMFT &  \centering  QCCMFT &  \centering exact/MCS &\\
    \hline
    Honeycomb   & \centering $\begin{array}{c} T_c\\ \Gamma_c\end{array} $ & \centering $\begin{array}{c} 3\\ 3\end{array} $  &  \centering $\begin{array}{c} 2.104 \\ 1.829 \end{array} $ 
    & \centering $\begin{array}{c} 2 \\ 2\end{array} $  & \centering $\begin{array}{c} 1.593 \\ 2.105\end{array} $   & \centering $\begin{array}{c}  1.519 \\ 2.132\end{array} $  & \\ \\
    Square       & \centering $\begin{array}{c} T_c\\ \Gamma_c\end{array} $ & \centering $\begin{array}{c} 4 \\ 4 \end{array} $    & \centering $\begin{array}{c} 3.090 \\ 2.751 \end{array} $ 
    & \centering $\begin{array}{c} 3 \\ 3\end{array} $   & \centering $\begin{array}{c} 2.362 \\ 2.956\end{array}$  &   \centering $\begin{array}{c} 2.269 \\ 3.044 \end{array} $   & \\ \\
   
    Cubic	 & \centering $\begin{array}{c} T_c\\ \Gamma_c\end{array} $ & \centering $\begin{array}{c} 6 \\ 6 \end{array} $    & \centering $\begin{array}{c} 5.073 \\ 4.704 \end{array} $ 
    & \centering $\begin{array}{c} 5 \\ 5\end{array} $   & \centering $\begin{array}{c} 4.763 \\ 5.198 \end{array} $ &\centering $\begin{array}{c} 4.511 \\ 5.158 \end{array} $  & \\
    \hline        
  \end{tabular}
  \label{table}
\end{table}


\section{Conclusion}
\label{conclusion}

In this article, we introduced a quantum correlated cluster mean-field method (QCCMFT), which had proven being capable of correctly describing classical and quantum phase transitions in some important spin systems. The reported method is an extension of the correlated cluster mean-field theory \cite{Yamamoto}, that was adapted here in order to treat many-body problems with quantum fluctuations. 
Our approach succeeds in better capturing quantum effects because it recognizes their influence on the spin-spin interactions, which is implemented through state superpositions determined self-consistently via the imaginary time spin self-interaction. 
As an application, the QCCMFT was used to analyse the transverse Ising model on three classes of spin lattices: honeycomb, square, and simple cubic. We found out that the QCCMFT improves the results for the spin correlations and takes into account more lattice geometric features when compared with other approximative methods, as for instance the MFT and EMFT. In particular, the QCCMFT enabled us to consider thermal and quantum effects in the same framework. As a result, this technique gives a very accurate location for the ferromagnetic-paramagnetic phase transition driven by thermal and quantum fluctuations. For instance, the quantum critical points obtained for the considered lattices present a difference of less than $3 \%$ when compared to those obtained using Monte Carlo simulations \cite{CMC_Deng}.
%

In summary, the simplicity of the approach and accurate results obtained here with the QCCMFT indicate that it shall be a valuable tool for investigations regarding quantum spin systems. It would be interesting to utilize this framework to deal with more complex systems, as for instance the geometrically frustrated ones. Another area in which this kind of tool can find application is for quantum correlation quantification in spin and other systems \cite{Vedral2004, Maziero2010, Maziero2012, Boette2015, Pasquale2010}, and its possible role for characterizing the so dubbed quantum criticality \cite{Sachdev2011, Merchant2014, Kinross2014}. It would be natural also using the QCCMFT to investigate the dynamics of many-body systems \cite{Kramer2015, Jin2016, Mendoza-Arenas2016}.

\begin{acknowledgements}
This work was supported by the Brazilian funding agencies: Conselho Nacional de Desenvolvimento Cient\'ifico e Tecnol\'ogico (CNPq), processes 441875/2014-9, 303496/2014-2, 474559/2013-0, 306720/2013-2, and 142382/2015-9, Instituto Nacional de Ci\^encia e Tecnologia de Informa\c{c}\~ao Qu\^antica (INCT-IQ), process 2008/57856-6, and Coordena\c{c}\~ao de Desenvolvimento de Pessoal de N\'{i}vel Superior (CAPES), process 6531/2014-08. JM gratefully acknowledges the hospitality of the Physics Institute and Laser Spectroscopy Group at the Universidad de la Rep\'{u}blica, Uruguay.
\end{acknowledgements}

\appendix \section{Imaginary time spin self-interaction}
\label{sec_appendix}

In this Appendix, the imaginary time spin self-interaction,
\begin{eqnarray}
 \bar{q} & = & \left\langle\frac{1}{\beta}\int_{0}^{\beta}d\tau  \sigma_{i}^{z}\sigma_{i}^{z}(\tau) \right\rangle_{\rho_{\beta}(H)} \nonumber \\
         & = & \mbox{Tr} \left(\frac{1}{\beta}\int_{0}^{\beta}d\tau \sigma_{i}^{z}\sigma_{i}^{z}(\tau)\rho_{\beta}(H)\right),
\end{eqnarray}
is evaluated. Above and hereafter we utilize $\sigma_{i}^{z}=\sigma_{i}^{z}(0)$. The Trotter formalism is used to deal with commutation relations present in the model. The ``integration variable'' $\tau$ is an imaginary time, with the associated time evolution operator leading to \cite{suzuki}: 
 $\sigma_{i}^{z}(\tau)= \mbox{e}^{-\tau H}\sigma_{i}^{z}\mbox{e}^{\tau H}.$

By considering 
a basis of eigenstates $|n\rangle$ for $H$ with corresponding energies $E_n$, 
we can evaluate  $\bar{q}$ as
\begin{eqnarray}
 \bar{q} & = & \frac{\displaystyle\sum_{n,m}\int_{0}^{\beta}d\tau \langle n|\sigma_{i}^{z}|m\rangle \langle m|\sigma_{i}^{z}|n\rangle \mbox{e}^{\tau(E_{n}-E_{m})}\mbox{e}^{-\beta E_{n}}}{\beta \sum_{n} \exp(-\beta E_{n})} \\
 & = & \frac{\sum_{n} \exp(-\beta E_{n}) |\langle n|\sigma_{i}^{z}|n\rangle |^2}{\sum_{n} \exp(-\beta E_{n})} \nonumber \\ 
 & & - \frac{\displaystyle\sum_{n\ne m} \frac{\exp(-\beta E_{m})-\exp(-\beta E_{n})}{E_m-E_n} |\langle m|\sigma_{i}^{z}|n\rangle|^2}{\beta\sum_{n} \exp(-\beta E_{n})}.
\label{qbar}
\end{eqnarray}

In particular, 
for the thermal state of a single spin with eigenstates $|\Phi^{+}\rangle$ and $|\Phi^{-}\rangle$ and corresponding energies $E_{+}$ and $E_{-}$, the spin self-interaction is given by:
\begin{eqnarray}
\bar{q} = \cos^{2}\theta +\sin^{2}\theta \frac{ \tanh\beta\Delta E/2}{\beta\Delta E/2},
\end{eqnarray}
with $\Delta E = E_{+} - E_{-}$. As our purpose here is to include quantum effects in mean-field approximations, we shall take the limit where thermal fluctuations are small compared to the energy gaps, i.e., $\beta\Delta E\rightarrow\infty$. In this limit, the second term in the right hand side of the last equation is null and $\bar{q}\approx\cos^{2}\theta$. Thus we shall use
 $\theta= \cos^{-1}\sqrt{\bar{q}}$ 
as motivation for our ansatz in the QCCMFT.

\bibliography{References}{}

\end{document}